\documentclass[preprint,showpacs,preprintnumbers,amsmath,amssymb]{revtex4}

\usepackage{graphicx}% Include figure files
\usepackage{dcolumn}% Align table columns on decimal point
\usepackage{bm}% bold math
 \usepackage[font=scriptsize,justification=raggedright, labelfont={bf}, margin=0cm]{caption}
\begin{document}

\title{Positive feedback produces broad distributions in maximum activation attained within a narrow time window in stochastic biochemical reactions}% Force line breaks with \\

\author{Jayajit Das\footnote{E-mail: das.70@osu.edu}}

\affiliation{Battelle Center For Mathematical Medicine, The Research Institute at the Nationwide ChildrenÕs Hospital and Departments of Pediatrics, Physics, and, Biophysics Graduate Program, The Ohio State University, 700 ChildrenÕs Drive, Columbus, OH 43205.}%

\date{\today}% It is always \today, today,
\begin{abstract}
How do single cell fate decisions induced by activation of key signaling proteins above threshold concentrations within a time interval are affected by stochastic fluctuations in biochemical reactions? We address this question using minimal models of stochastic chemical reactions commonly found in cell signaling and gene regulatory systems. Employing exact solutions and semi-analytical methods we calculate distributions of the maximum value ($N$) of activated species concentrations ($P_{max}(N)$) and the time ($t$) taken to reach the maximum value ($P_{max}(t)$) within a time window in the minimal models. We find, the presence of positive feedback interactions make $P_{max}(N)$ more spread out with a higher ``peakedness" in $P_{max}(t)$. Thus positive feedback interactions may help single cells to respond sensitively to a stimulus when cell decision processes require upregulation of activated forms of key proteins to a threshold number within a time window. 
\end{abstract}
%\pacs{72.25.Dc, 72.25.Hg, 72.25.Mk}% PACS, the Physics and Astronomy
\maketitle
\section{Introduction}

Decisions made at the single cell level enable organisms to respond to changes in the local environment.  Such decisions are usually processed upon upregulation of specific proteins, transcription factors, or soluble molecules that help cells to communicate with each other. These activation events often require concentrations of few key proteins to reach a threshold level within a time window.  Examples of such responses include, activation of immune cells such as T cells triggered by a threshold number of pathogenic peptides \cite{DAS}, all or none maturation of oocytes in the frog {\it Xenopus laevis} induced by different concentrations of progesterone \cite{FERR1}, or  switch like activation of $Lac$ genes regulating lactose metabolism produced by a threshold concentration of stimulus in {\it E. coli} \cite{COLLINS}. 

 However, every cell in a cell population interacting with stimuli possesses unique temporal profiles of concentrations of activated signaling molecules or genes. This cell to cell variability in the kinetics occurs due to the inherent stochastic nature of associated biochemical processes (or intrinsic noise) \cite{DELBRUCK, ARKIN,GARDN,SIGGIA} and variations in expression levels of genes and proteins (or extrinsic noise) \cite{SIGGIA}. Therefore, the threshold for activation for a specific signaling molecule and the time window within which the signaling molecule should be activated to influence cell functions can change from cell to cell \cite{FERR1}. How do nonlinearities commonly found in biochemical signaling networks, such as positive feedbacks, help cells to respond to these variations? We address this question in the article, in particular, we investigate the role of positive feedback interactions which are often responsible for producing all or none responses in signaling or gene regulatory kinetics.  We use a minimal model for a linear and a positive feedback interaction in a simple chemical reaction describing activation of a single chemical species representing a key signaling protein or a gene. Since many positive feedback interactions \cite{DAS,FERR1,WEINBERGER}can be reduced to this form the results from the model will be relevant for a wide range of biological systems. 
 
We consider a minimal biochemical process, $C\mathop{\leftrightharpoons} C^*$, describing production and deactivation of the activated species  $C^*$ which needs to reach a threshold concentration (say $N$) in a time interval $[0,T]$ in order to mediate a functional response. Due to the stochastic fluctuations in the kinetics, the threshold concentration of $C^*$ could occur at different times (Fig. 1) or even stay below the threshold level in the time window. Therefore, knowing the distribution of the number ($n$) of molecules of $C^*$ at a time $T$ will not reveal if the concentration of $C^*$ attained the threshold level at an earlier time. However, knowledge of the joint distribution of the maximum number ($N$) of $C^*$ and the time $t$ ($0\le t \le T$) when this value was attained in a temporal profile describing the kinetics of $C^*$ in a single cell will inform us if the cell was able to cross the threshold in the time interval $[0,T]$. Such distributions are regularly dealt with in Extreme Value Theory (EVT) where extreme value distributions for identically distributed independent random variables have been studied extensively \cite{GUMB}.  Analysis of extreme value distributions for correlated random variables has been a topic of intense research in the recent years due to its application in physics\cite{MEZARD,MAJUM1, MAJUM2, RACZ, REDNER},  climate science \cite{KATZ}, finance\cite{FINANCE}, and, population\cite{GILLESPIE1,ORR} and cell biology\cite{CHAK}. Application of such distributions in stochastic biochemical reaction kinetics has been initiated only recently\cite{KARDAR}. Interestingly, it has been found that for strongly correlated random variables in different types of random walks or fluctuating interfaces extreme value distributions can display simple one parameter scaling behavior \cite{MAJUM1, MAJUM2}.

We solve the Master Equation associated with the minimal model and calculate the joint probability distribution for $C^*$ attaining a maximum value $N$ at time $t$ in the time interval $[0,T]$ exactly analytically and semi-analytically.  We show that when the system is far from the steady state, in the presence of the feedback reaction, the distribution of the maximum value $N$ over the time interval $[0,T]$ is  spread out over a broader range of $N$ compared to the linear model. In contrast, the distribution of the time $t$ when the maximum value occurred is much narrowly distributed in the presence of the feedback. This suggests that feedback interactions can help single cells to respond sensitively to weak stimulus with a well-defined response time even in the face of stochastic fluctuations. 

\section{Results}

\subsection{Irreversible kinetics}
 
In order to understand the role of stochastic fluctuations in affecting the distribution of maximum value of $C^*$, it will be instructive to study the deterministic mass action kinetics for the concentration of $C^*$ (or $[C^*]$) in the reaction,  $C\mathop{\leftrightharpoons} C^*$, which is described by, 
\begin{equation}
d[C^*]/dt=(k_1+k_p[C^*])[C]-k_{-1}[C^*] \, . 
\label{eqn:rate}
\end{equation}
The total concentration, $C_0=[C]+[C^*]$, is always fixed, and, the rates $k_1$ and $k_p$ determine timescales for production of $C^*$ from $C$ via a linear first order reaction, and, a second order reaction representing a positive feedback, respectively. $C^*$ is converted back to $C$ with a rate, $k_{-1}$. These time scales in a biological network can be regulated by the strength of a stimulus that results in generation of $C^*$, e.g., a weaker (or stronger) stimulus would give rise to longer (or shorter) time scales for production of $C^*$. The rate equation contains a single stable fixed point, and, thus starting with any initial concentration, $[C^*]$ monotonically reaches a steady state determined by the rate constants, and, $C_0$. Consequently, if a reaction initiated with  a concentration $[C^*(t=0)]<[C^*(t\rightarrow \infty)]$ is followed until $t=T$, the maximum value of  $[C^*]$, uniquely determined by the rate constants, $C_0$, and $T$, is reached at $t=T$.  However, in the presence of intrinsic stochastic noise fluctuations, the maximum value of the concentration of $C^*$ or the time when it is attained will vary in each stochastic `trajectory' (Fig. 1), where every trajectory represents activation of $C^*$ in a single cell.  In this situation,  $P(n,t|m,0)$, the conditional probability of having $n$ number of molecules of the $C^*$ species at any time $t$ starting with a distribution $P(m,0)$ at $t=0$, follows the Master Equation, 
\begin{eqnarray}
 \frac{\partial P(n,t|m,0)}{\partial t}  &  & =  ({N_0} - n + 1)({k_1} + {k_p}(n - 1))P(n - 1,t|m,0)  + {k_{ - 1}}(n + 1)P(n + 1,t|m,0)  \nonumber \\  
 & & - ({k_1}({N_0} - n) + {k_{ - 1}}n + {k_p}n(N_0-n))P(n,t|m,0) \, ,
 \label{eqn:master}
\end{eqnarray}
where, $N_0$ denotes the total number of molecules of $C$ and $C^*$ species. The distribution of the maximum number ($N$) of  $C^*$ molecules and the time when the maximum was reached in a time interval $[0,T]$ can be calculated by solving of the above Master equation and using the renewal equation \cite{GOEL,HONER} , 
\begin{equation}
P(\left. {n,t} \right|m,0) = {Q_{N }}(\left. {n,t} \right|m,0) + \int\limits_0^t {dt'\;} {F_{N }}(\left. {t'} \right|m,0)P(\left. {n,t} \right|N,t'). 
\end{equation}

${Q_{N}}(\left. {n,t} \right|m,0)$ describes the probability of having $n$ molecules of $C^*$ species at time $t$, when an absorbing boundary condition, ${Q_{N}}(\left. {n,t} \right|m,0) = 0$ for $n \ge N$, is imposed.  ${F_{N}}(\left. t \right|m,0)$ denotes the probability of arriving at the state $n=N$ for the first time at time $t$. If the time variable is Laplace transformed in the renewal equation, then ${F_N}(\left. s \right|m,0)$ is related to $P(\left. {N,s} \right|m,0)$ simply by, ${F_N}(\left. s \right|m,0) = P(\left. {N,s} \right|m,0)/P(\left. {N,s} \right|N,0)$. The unnormalized joint probability distribution for attaining a maximum value $N$ at time $t$ in the time interval $[0,T]$ is then given by, 

\begin{equation}
{E_N}(T,t|m,0) = \sum\limits_{n = 0}^N {{F_N}(\left. t \right|m,0){Q_{N + 1}}(\left. {n,T} \right|N,t)}\,.
\end{equation}
We then calculate the unnormalized distribution of the maximum value $N$ over the time interval $[0,T]$, $i.e.$, 
\begin{equation}
P_{max}(N,T)=\int_0^{T} dt\, {E_N}(T,t |m,0)
\end{equation}
and the unnormalized distribution of the time $t$ when the maximum value occurred given by, 
\begin{equation}
P_{max}(t,T)=\sum_{N=m+1}^{N_0}\, {E_N}(T,t |m,0) \, .
\end{equation} 

We first consider the case where the production of $C^*$ occurs irreversibly, $i.e.,$ $k_{-1}=0$. In this limit, ${E_N}(T,t,|m,0)$ can be evaluated analytically as calculations simplify due to the following relations: $P(\left. {N,T} \right|m,0) =0$ for $m>N$, thus, ${Q_{N + 1}}(\left. {N,T} \right|m,0) = P(\left. {N,T} \right|m,0)\quad $ for $m \le N$. Consequently,  the joint probability distribution can be expressed as, ${E_N}(T,t,|m,0) = {F_N}(\left. t \right|m,0)P(N,\left. T \right|N,t)$. This essentially implies that the probability of having a maximum value $N$ at time $t$ is the probability the state $n=N$ was attained at time $t$ for the first time and then no reaction occurred in the time interval $T-t$. Next we calculate these distributions for the linear and the feedback models by solving Eq. \ref{eqn:master} for $k_{-1}=0$.

In the absence of the positive feedback ($k_p=0$), the exact solution of the Master Equation in Eq.\, \ref{eqn:master} yields, $P(\left. {N,t} \right|m,0){ = ^{{N_0} - m}}{C_{{N_0} - N}}\,{e^{ - ({N_0} - N){k_1}t}}{(1 - {e^{ - {k_1}t}})^{N - m}}$, where, $P(m,0)=\delta_{m,0}$. The first passage time distribution is given by, ${F_N}(\left. t \right|m)=({N_0} - N + 1){k_1}P(\left. {N - 1,t} \right|m,0)$, therefore, ${E_N}(T,t,|m,0) = k_1({N_0} - N + 1)P(\left. {N - 1,t} \right|m,0)P(N,\left. T \right|N,t)$. Thus, $P_{max}(N,T)= P(\left. {N,T} \right|m,0)$, and, ${P_{\max }}(T,t)=({N_0} - m){e^{ - {k_1}t}}{\left( {1 + {e^{ - {k_1}T}} - {e^{ - {k_1}t}}} \right)^{{N_0} - (m + 1)}}$ (see \cite{SUPPL} for additional details). In the presence of the feedback, Eq.\, \ref{eqn:master} can be solved exactly by Laplace transforming the time variable. We consider the `feedback only' ($k_1=0$ and $k_p\ne 0$) case to exclusively interrogate the role of the positive feedback. The exact solution for the time dependent probability distribution for $m=1$ is given by, 
\begin{equation}
P(\left. {N,s} \right|1,0)=k_p^{N - 1}(N - 1)!^2 \,^{{N_0} - 1}C_{N_0 - N}\prod\limits_{r = 1}^N {\frac{1}{{(s + {k_p}r({N_0} - r))}}}
\end{equation}
The calculation of the inverse Laplace transformation of the above equation is tedious but straightforward, and, since the poles of $P(\left. {N,s} \right|1,0)$ at two different values of $r$ can be equal, the probability distribution contains terms which are product of linear and exponential functions of $t$ (details in \cite{SUPPL}). The first passage time distribution for this case is given by, ${F_N}(\left. t \right|m)={k_p}(N - 1)({N_0} - (N - 1))P(\left. {N - 1,t} \right|1,0)$, therefore, ${E_N}(T,t,| 1,0)={k_p}(N - 1)({N_0} - (N - 1))P(\left. {N - 1,t} \right|1,0){e^{ - {k_p}N({N_0} - N)(T - t)}}$. As in the linear model we find, $P_{max}(N,T)= P(\left. {N,T} \right|1,0)$. However, $P_{max}(t,T)$ does not possess a simple expression as the linear model. The shapes of the distributions, $P_{max}(N,T)$ and $P_{max}(t,T)$, depend on $N_0$ and the dimensionless variable, $\tau=k_pT$ (or $k_1T$ for the linear model). In order to compare the distributions for the pure feedback and the linear models, we chose an end time $T$, where the average number of $C^*$ molecules was the same for both the models. Fig. 2a shows the maximum value is distributed more evenly across different numbers of molecules of $C^*$ in the presence of the feedback compared to the linear model, in contrast,  $P_{max}(t,T)$ (Fig. 2b) is more sharply peaked for the feedback model, indicating that once the first molecules of  $C^*$ are produced the positive feedback leads to fast production of $C^*$ molecules giving rise to a peak at $t=T$. The variation (inset, Fig. 2a) of the Fano Factor, $f= (\left\langle {{N^2}} \right\rangle  - {\left\langle N \right\rangle ^2})/\left\langle N \right\rangle$, which quantifies if a distribution is broader than a Poisson distribution (where, $f=1$), with $\left\langle {N} \right\rangle$ at different times shows that $P_{max}(N,T)$ is more spread out for the pure feedback model as long as the system is away from the steady state. As the system approaches the steady state, due to the irreversibility in the reactions, all the $C$ molecules are converted into $C^*$, $i.e.$ , $P(N,t\rightarrow \infty | m,0) \rightarrow \delta_{N,N_0}$, and then $f$ decreases and become comparable for both the models. We used Kurtosis ($K$) defined as, $K=\mu_4/\sigma^4-3$, where, $\mu_4$ and $\sigma^2$ denote the 4th cumulant and the variance, respectively, to quantify the ``peakedness" and the presence of ``heavy tails" \cite{DECAR} in $P_{max}(t,T)$ compared to a Gaussian distribution ($K=0$). The pure feedback model produces much larger values of $K$ at different values of $\left\langle {N} \right\rangle$ compared to the linear model (inset, Fig. 2b) indicating higher ``peakedness" of the distributions in the presence of the feedback. 

The above results become evident at the limit, ${N_0} \to \infty$, ${k_1}{N_0} \to {\tilde k_1}$ and ${k_p}{N_0} \to{\tilde k_p}$. Then the probability distribution, $P(\left. {n,t} \right|m,0)$, follows a simple functional form for both the models. In the linear model, 
\begin{equation}
{P_{\max }}(N,T) = {\tilde k_1}^{N - 1 - m}e^{ - {\tilde k}_1T}( {{{\tilde k}_1}T})^{N - m}/(N - m)!
\end{equation}
 and 
\begin{equation}
{P_{\max }}(T,t) = {e^{ - {{\tilde k}_1}(T - t)}} \, ,
\label{eqn:tprobl}
\end{equation} 
whereas, in the model with pure feedback\cite{DELBRUCK}, 
\begin{equation}
P_{\max }(N,T)=m(m + 1) \cdots (N - 1)(1 - e^{ - {{\tilde k}_p}T})^{N - m}e^{ - {{\tilde k}_p}mT}/(N - m)! 
\end{equation}
and 
\begin{equation}
P_{\max }(T,t) = me^{ - m{{\tilde k}_p}T}e^{ - {{\tilde k}_p}(T - t)}[ {1 - (1 - {e^{ - {{\tilde k}_p}t}})e^{ - {{\tilde k}_p}(T - t)}}]^{-(m + 1)} \, .
\label{eqn:tprobf}
\end{equation}
Calculation of the Fano factor ($f$) for $P_{max}(N,T)$ shows that, $f = {\tilde k_1}T/(m + {\tilde k_1}T)<1$ and $f=e^{{\tilde k}_pT}-1$ for the linear and the feedback model, respectively; clearly demonstrating that $P_{max}(N,T)$ contains more variation in $N$ for the feedback model for $T>1/{\tilde k}_p$. On the other hand, the denominator in $P_{max}(t,T)$ for the feedback model decreases as $t$ approaches $T$ making the distribution more sharpely peaked at $t=T$ compared to the linear model.  These results provide the following physical understanding. For the feedback model, the probability ($\propto e^{-n\tilde{k}_p\Delta t}$) to remain in the state $n$ for a time interval $\Delta t$ decreases with $n$, whereas, this probability ($\propto e^{-\tilde{k}_1\Delta t}$) is independent of $n$ for the linear model. Therefore, in the feedback model, the states with smaller values of $n$ spend large fraction of their time initially in waiting for the first reactions to occur, and then at times closer to the end time $T$, as these states move to larger values of $n$, the next reactions take place in rapid successions. This produces a large range of values of maximum values $N$ in the time interval $[0,T]$ but a sharp peak in $P_{max}(t,T)$. In contrast, in the linear model a state with $n$ molecules moves to the next higher state  ($n+1$ molecules) with a constant rate, so at $t=T$ all the states reside close to the average value of $C^*$. In addition, since the waiting time for next reaction to occur does not depend on $n$, $P_{max}(t,T)$ is more spread out. We expect these features to persist even in the presence of a non-zero de-activation rate, as long as, the time scales for the feedback reactions are smaller than the de-activation time scale. These results suggest the following biological significance. During early signaling events, if the rate of de-activation is slower than that of activation, then in time scales smaller than that of deactivation the presence positive feedback interactions can help cells to achieve a wide range of activation within a narrow response time even in the presence of stochastic fluctuations.  In the following section we will show the main features of the distributions $P_{max}(N,T)$ and $P_{max}(t,T)$ persist even in the presence of a non-zero de-activation rate.

\subsection{Reversible kinetics}

In the presence of a non vanishing rate of de-activation (i.e., $k_{-1} \ne 0$), the number of $C^*$ molecules can decrease after attaining the maximum value, therefore, the simple relationships between the $P(n,t|m,0)$ and the maximum value distributions as in the previous section no longer hold. Moreover, it becomes difficult to analytically solve the Master Equation exactly when the positive feedback is present. Therefore, we calculated the maximum value distributions semi-analytically. We briefly outline the method here, the details of the calculations are shown in \cite{SUPPL}. The Master Equation in Eq.\,\ref{eqn:master} can be cast as an operator equation \cite{HONER} described by, 
\begin{equation}
{\partial \left| {P(t)} \right\rangle }/{{\partial t}} = L\left| {P(t)} \right\rangle \, ,
\end{equation}
where, $\langle n | P(t)\rangle  = P(n,t|m,0)$ and ${(L)_{nn'}} = ({N_0} - n')({k_1} + {k_p}n'){\delta _{n',\;n - 1}} + {k_{ - 1}}n'{\delta _{n',\;n + 1}} - ({k_1}({N_0} - n) + {k_{ - 1}}n + {k_p}nn'){\delta _{n',\;n}}$. We solve the above equation by numerically evaluating the right ($\left| {{R_r}} \right\rangle $) and left ($\left\langle {{L_r}} \right|$) eigenvectors, and the eigenvalues ($\{ \lambda_r\}$) of the operator, $L$. 
The solution of the Master equation then can be written as, $\left\langle n| {P(t)} \right\rangle  = P(n,t|m,0)=\langle n | {e^{L\,t}}\left| {P(0)} \right\rangle  = \sum\limits_{r = 0}^{{N_0}} {{e^{{\lambda _r}\,t}}} {a_r}(0)\left\langle n | {{R_r}} \right\rangle  = \sum\limits_{r = 0}^{{N_0}} {} {a_r}(t) {{R_{rn}}}$, where, $\left| {P(0)} \right\rangle $ describes the probability distribution at $t=0$, and, ${R_{r{\kern 1pt} n}} = \left\langle {n\left| {{R_r}} \right\rangle } \right.$. $\{ {a_n}(0)\} $ is calculated from the initial condition. The same scheme is used to calculate the probability distribution, ${Q_N}(\left. {n,t} \right|m,0)$ using ${Q_N}(\left. {n,t} \right|m,0) = \sum\limits_{r = 0}^{{N_0}} {{e^{\lambda _{_r}^{(N)}t}}} a_r^{(N)}(0)R_{r\,n}^{(N)}$, where, $\{ \lambda _{_r}^{(N)} \}$ and $\{R_{r\,n}^{(N)} \}$ are the eigenvalues and eigenvectors of $L$ with an absorbing boundary condition at $n=N$, respectively. We define the survival probability, ${S_N}(\left. t \right|m,0) = \sum\limits_{n = 0}^N {{Q_N}} (\left. {n,t} \right|m,0)$, which can be used to evaluate the first passage time distribution, ${F_N}(\left. t \right|m,0)$ from ${F_N}(\left. t \right|m,0) =  -{\partial {S_N}}/{\partial t}$. The maximum value distribution functions are then calculated using the following equations, \begin{eqnarray}
P_{\max }(N,T) = \int\limits_0^T {dt\,{F_N}} (\left. t \right|m,0){S_{N + 1}}(\left. {T - t} \right|N,0)\, ,
\end{eqnarray}
and,
\begin{equation}
P_{\max }(T,t) = \sum\limits_{N = m + 1}^{{N_0}} {{F_N}(\left. t \right|m,0){S_{N + 1}}(\left. {T - t} \right|N,0)}  \, .
\label{eqn:ptmaxop}
\end{equation}

The shape of the probability distributions, $P(n,t|m,0)$ and $E_N(t,T|m,0)$ depend on the dimensionless parameters, $k_p t$, $k_1/k_p$ and $k_{-1}/k_p$ for the feedback model, and, $k_1t$ and $k_{-1}/k_1$ for the linear model. We varied $k_{-1}$ as well as $t$ and $T$ in the models to investigate the effect of the de-activation rate on the above distributions. We kept $k_1/k_p$ fixed to a small non-zero value ($0.01$) with two goals in mind: (i) prevent the $n=0$ state from becoming an absorbing state, (ii) to exclusively study the effect of the feedback; when $k_1/k_p \gg 1$, the feedback model starts behaving like the linear model. The presence of a non-zero de-activation rate makes $P_{max}(N,T)$ different than $P(n,T|m,0)$ (Fig. 3a), since $P(n,t|m,0)$ no longer vanishes for $m>n$. The non zero deactivation rate can generate bimodal distributions for $P(n,T|m,0)$ (Fig. 3a) in the feedback model, which is purely generated due to stochastic fluctuations since the deterministic rate equation (Eq. \,\ref{eqn:rate}) does not possess any bistability. $P_{max}(N,T)$ can also show a bimodal distribution as $P(n,T|m,0)$. However, the peaks in $P_{max}(N,T)$ occur at larger values of $N$ compared to the values of $n$ where $P(n,T|m,0)$ is peaked. This effect is produced by stochastic trajectories that attained the maximum value $n=N$  at times earlier than $T$. Since $C^*$ in the linear and the feedback models can attain the state $n=N_0$ with a non vanishing probability, $P_{max}(N,T) \rightarrow \delta_{N,\, N_0}$ as $T\rightarrow \infty$ in both the models, making the distributions similar for both the models at long times. Therefore, the bimodal distribution in $P_{max}(N,T)$ for the feedback model will be transient. These results can have implications for transient bimodal distributions observed in experiments when the underlying signaling network predicts steady state bimodal distributions. In such experiments, activation of the cells could be caused by a key signaling protein crossing the activation threshold for the first time, consequently, the distribution of activated cells will be represented more appropriately by $P_{max}(N,T)$ instead of $P(n,T|m,0)$.  Next we calculated $P_{max}(t,T)$ using Eq. \ref{eqn:ptmaxop}. As in the irreversible case, the distribution, $P_{max}(t,T)$, shows a higher peakedness  for the feedback model as compared to the linear model (Fig. 3b). The larger values of the Fano factor ($f$) for the distribution $P_{max}(N,T)$  for a range of de-activation rates (Fig. 4a and 4c) and end times $T$ demonstrate that $P_{max}(N,T)$ continues to have more variance for the feedback model compared to the linear model as long as the system is away from the steady state.  The peakedness in $P_{max}(t,T)$ was characterized by the kurtosis, $K$, as in the previous section.  The feedback model produces positive and substantially larger values of $K$  than that for the linear model (Fig. 4b and 4d), where, for most of the parameter values, $K$ is negative, indicating a flatter distribution compared to a Gaussian distribution.  Therefore, the above results demonstrate that even in the presence of de-activation of signaling molecules, the presence of the positive feedback prepares the single cells to respond to a wide range of activation thresholds in a well defined time window in a noisy environment.

\section{Conclusion}
We have studied how commonly found nonlinear biochemical processes in cell signaling and gene regulatory networks  such as positive feedbacks influence single cell decision processes in the presence of stochastic fluctuations when these decisions are regulated by key proteins attaining threshold concentrations within a time window. We analyzed the joint probability distribution of the maximum value of concentration of a molecular species and the time when the maximum concentration was reached instead of the probability distribution of the concentration of the molecular species at any time, as the later distribution does not contain information regarding if the molecular species attained the threshold concentration at an earlier time. We calculated the maximum value distributions exactly and semi-analytically in minimal models that can effectively describe linear and positive feedback interactions in biochemical reactions found in a wide range of cell signaling networks. In particular, we investigated the role of positive feedback interactions in affecting the shape of the maximum value distributions. We find that in the presence of a positive feedback interaction the maximum values of concentrations of the activated species are distributed more broadly compared to the linear model when the system is away from the steady state. However, the positive feedback produces a narrower distribution in the time when the maximum activation was achieved.  Therefore, a positive feedback interaction, even in situations when stochastic fluctuations dominate signaling kinetics,  provides cells the ability to respond to situations when specific cellular proteins need to attain a wide range of threshold concentrations within a narrow time window to influence cell decision processes. This property of the positive feedback could play an important role when a cell population has to sensitively respond to a weak stimulus. The biological significance of the presence of a ``heavy tail" at time scales smaller than the most probable time scale in $P_{max}(t,T)$ in the presence of positive feedback interactions is less evident. Perhaps, the positive feedback helps create a small reservoir of cells that can react to very weak stimuli with a range of relatively smaller response times when the majority of the cells that are destined to respond at a much biologically irrelevant longer time scale. The probability distributions calculated from the solutions of the Master equations for the minimal models indicate presence of multiple time scales in the system. It will be interesting to see if this can produce multi scaling behavior \cite{KADAN} in extreme value distributions in such chemical reactions in general. Such examples will be qualitatively different than extreme value distributions in standard Brownian motion \cite{MAJUM2} or models of fluctuating interfaces \cite{MAJUM1} which display single parameter scaling.  In addition, the minimal models studied here are embedded in larger biological networks which ultimately determine cell fate responses, therefore, it will be important to investigate the behavior of maximal value distributions when the minimal models are embedded in a larger network. 

\vspace{5mm}
%\noindent{{\bf Acknowledgments}}
This work was funded by the Research Institute at the Nationwide Children's Hospital and a grant (1R56AI090115-01A1) from the NIH. I thank C. Jayaprakash and M. Kardar for discussions, S. Mukherjee for  help with LAPACK routines, and, anonymous reviewers for their helpful comments.
%Create the reference section using BibTeX:
%\noindent{{\bf References}}

\newpage
{\bf Figure Captions}

\begin{figure}[h] %  figure placement: here, top, bottom, or page
   \centering
 \includegraphics[width=4in]{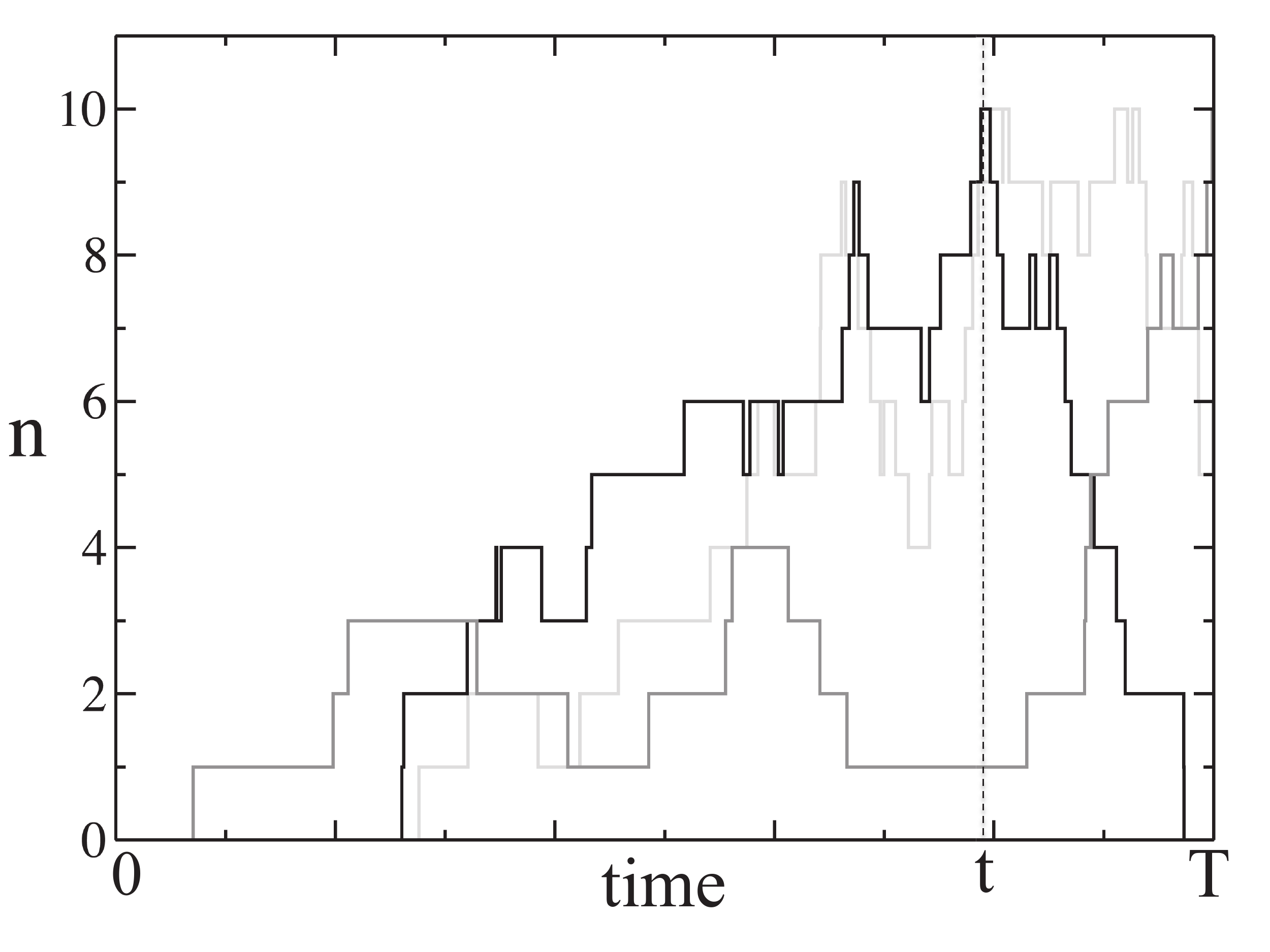} 
   \caption{Shows three different temporal profiles of the number ($n$) of  $C^*$ molecules that attain a maximum value $N=10$ in the time interval $[0,T]$. Each profile, representing activation of $C^*$ in a single cell, shows cell to cell variation of the time $t$ when the maximum value was reached. The stochastic trajectories were obtained by performing a Gillespie simulation \cite{GILLESPIE,SSC} of the chemical reaction $C\mathop{\leftrightharpoons} C^*$, where the system started with no $C^*$ molecules at $t=0$ for the parameters (described in Eq.\,\ref{eqn:master}) $k_1=0.001$, $k_p=0.01$,  $k_{-1}=0.5$, and $N_0=50$. }
   \label{fig:traj}
\end{figure}

\begin{figure}[h] %  figure placement: here, top, bottom, or page
   \centering 
 \includegraphics[width=5in]{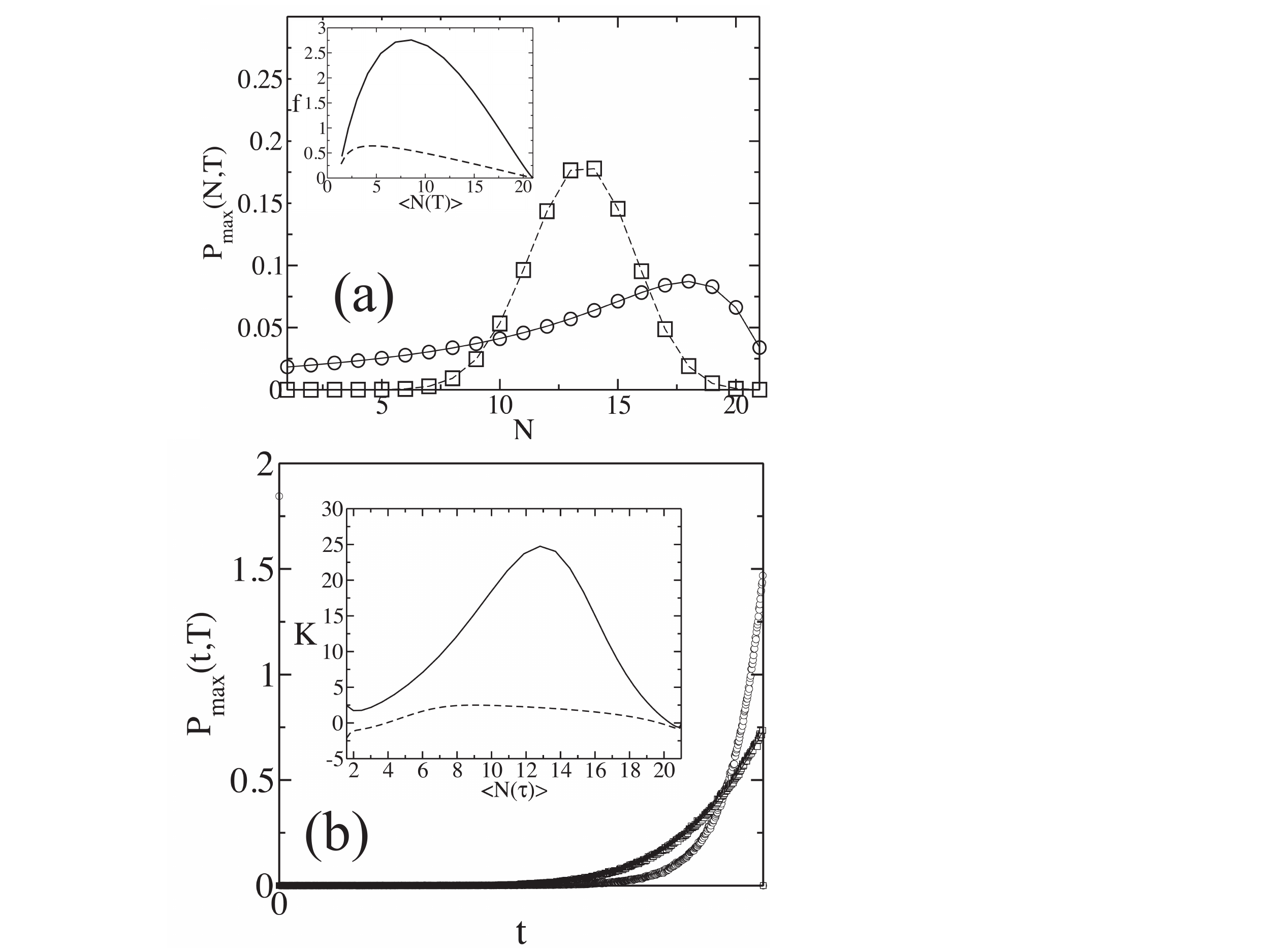} 
   \caption{Comparison between the feedback and the linear model when $k_{-1}=0$, $N_0=21$, and $P(m,0)=\delta_{m,1}$. The results from the exact solutions (solid lines:feedback model; dashed lines:linear model) are compared with Gillespie simulations \cite{GILLESPIE, SSC} of the reactions for the feedback ($\circ$) and the linear ($\Box$) model. Data from the Gillespie simulations are averaged over $10^6$ stochastic trajectories.  (a) Shows $P_{max}(N,T)$ at $T=10$  for the feedback ($k_p=0.01$) and the linear ($k_1=0.9699$) models where both the models produce the same values of  $\langle N(T)\rangle$. (Inset) Variation of the Fano factor $f$ with $\langle N(T)\rangle $ calculated from $P_{max}(N,T)$ at different times for the same parameter values as in (a). The feedback model (solid line) displays larger values of $f$ as compared to the linear model (dashed line) before the systems reach the steady states. (b)  Shows the distributions $P_{max}(t,T)$ vs $t$ at $T=10$ for the linear and the feedback models. The other parameters are the same as in (a). The point at $t=0$ shows the probability for the $n=1$ state to remain in the same state until $t=T$. (Inset) Shows variation of Kurtosis $K$ with $\langle N(T)\rangle $ as $T$ is increased. The feedback model displays larger values of $K$ indicating presence of larger peaks and heavy tails in $P_{max}(t,T)$.}
 \label{fig:irrev}
\end{figure}

\begin{figure}[ht] %  figure placement: here, top, bottom, or page
   \centering
      \includegraphics[width=5.0in]{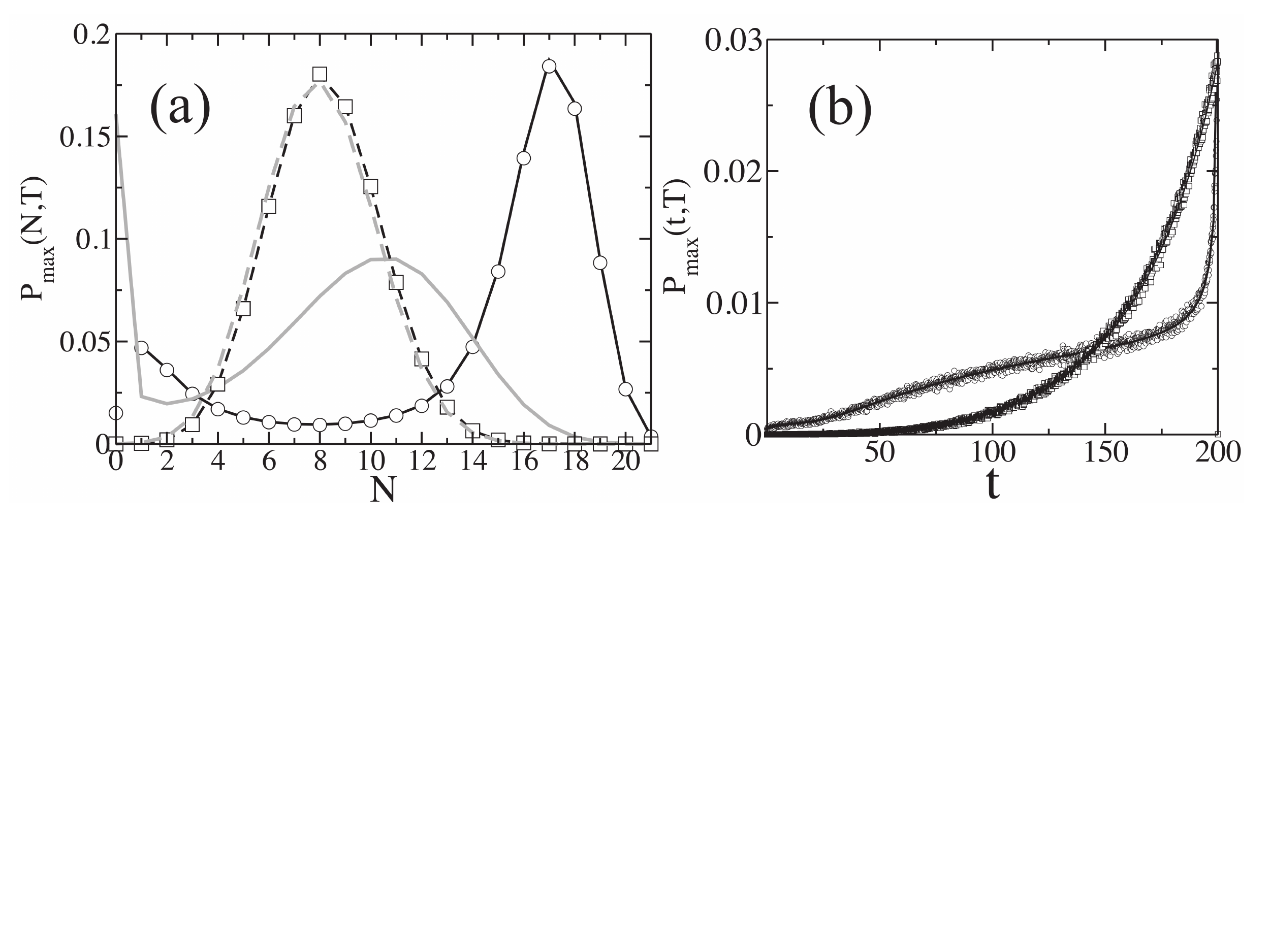} 
   \caption{Comparison of the distribution functions $P_{max}(N,T)$ and $P_{max}(T,t)$ between the feedback and the linear models for non zero de-activation rates ($k_{-1} \ne 0$). The results are shown for $N_0=21$ and an initial distribution of $C^*$ given by $P(m,0)=\delta_{m,0}$ for both the models. (a) Shows $P_{max}(N,T)$ at $T=200$ for the feedback ($k_p=0.01\,, k_{-1}=0.1$ and $k_1=0.001$) and the linear model ($k_{-1}=0.01$ and $k_1=0.00269$) where both the models produce the same values of  $\langle N(T)\rangle$. Data from Gillespie  simulations \cite{GILLESPIE, SSC} averaged over $10^6$ stochastic trajectories for the feedback ($\circ$) and the linear ($\Box$) models are compared with the semi-analytical calculations (dark solid lines: feedback; dark dashed lines:linear). $P(n,T|m,0)$, calculated using the semi-analytical scheme for the feedback (gray solid line) and the linear (gray dashed line) models, are compared with $P_{max}(N,T)$. (b) Variation of $P_{max}(t,T)$ with $t$ at $T=200$ for the linear and feedback models. The results from the semi-analytical calculations are shown in dark solid and dashed lines for the feedback and the linear models, respectively. $P_{max}(0,T)$ for the feedback model showing the probability of the system to remain at the initial state ($m=0$) until time $t=T$ is finite and is not shown on the graph. Data from Gillespie simulations, shown using the same visualization scheme as in (a), are compared with the semi-analytical calculations. All the parameters used for the calculations are the same as in (a).}
\label{fig:rev}
\end{figure}
   
\begin{figure}[ht] %  figure placement: here, top, bottom, or page
 \centering
  \includegraphics[width=6.0in]{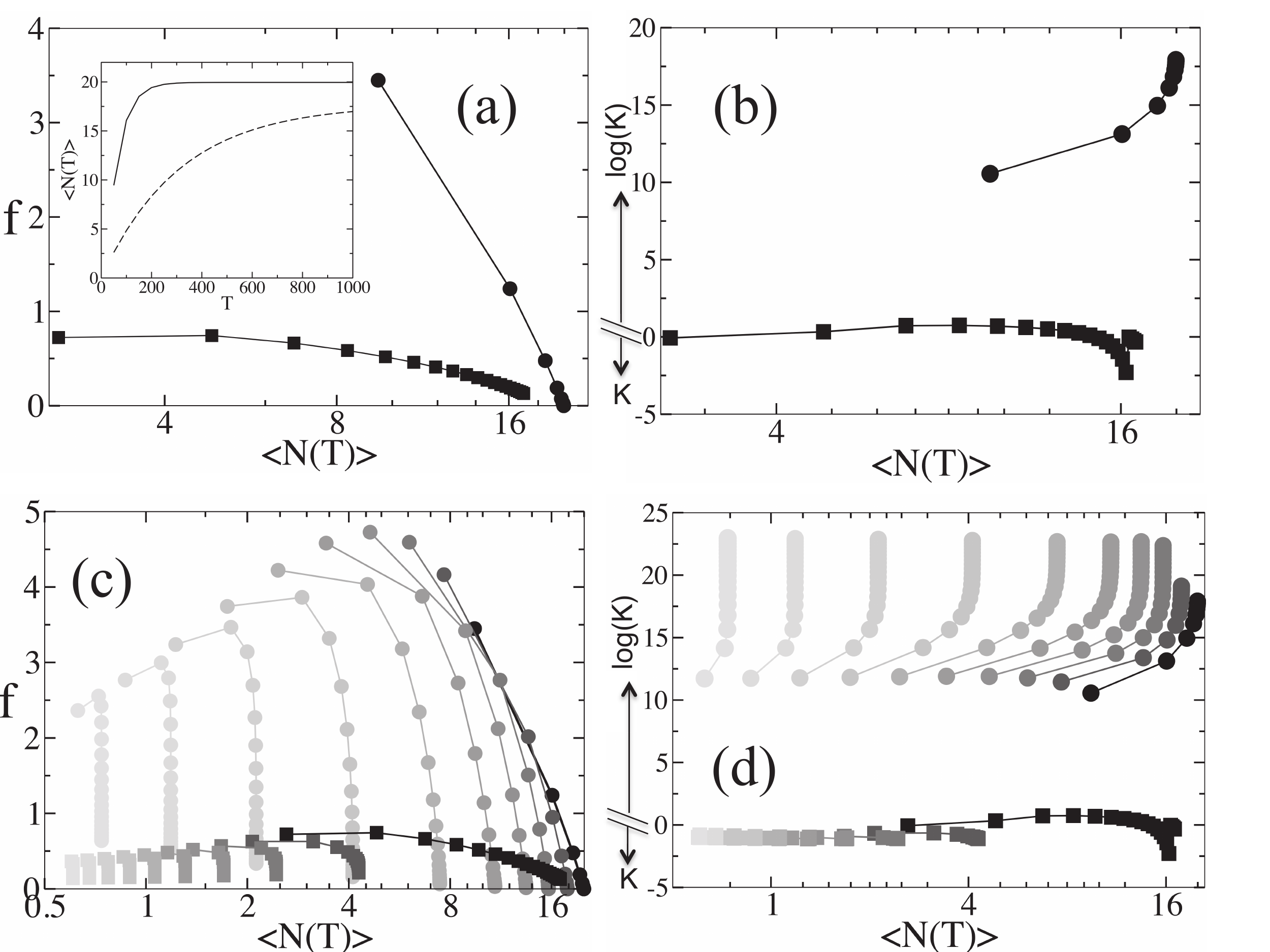} 

\caption{ Comparison of the shapes of the distributions $P_{max}(N,T)$ and $P_{max}(T,t)$, characterized by Fano factor, $f$, and, Kurtosis, $K$, respectively, between the feedback and the linear  models for non zero de-activation rates ($k_{-1} \ne 0$).  The results shown are calculated using the semi-analytical scheme for $N_0=21$ and an initial distribution of $C^*$ given by $P(m,0)=\delta_{m,0}$.  The results from the feedback  and the linear models are shown with filled circles and filled squares, respectively.
   (a) Shows variation of  the Fano factor, $f$, with $\langle N(T)\rangle$ calculated using the semi-analytical scheme described in the main text at different values of $k_{-1}$  as $T$ is increased uniformly (interval of $50$) from $50$ to $1000$ for the feedback and the linear model. The rates used for the feedback model are given by, $k_p=0.01$, $k_1=0.001$, and, $k_{-1}=0.01$. The rates for the linear model are set at $k_1=0.00269$ and $k_{-1}=0.0005$. 
The figure shows that the values of $f$ are substantially larger in the feedback model than the linear model for the same values of $\langle N(T)\rangle$ as long as the system is away from the steady state. (Inset) The average value of $C^*$, $\langle N(T)\rangle$, calculated for the same parameters used in (a), increases with $T$ until the system reaches the steady state for both the feedback (solid line) and the linear (dashed line) model. (b) Variation of $K$ with $\langle N(T)\rangle$ for the feedback and the linear model are displayed using the same visualization scheme and parameters as in (a). The feedback and the linear models produced large +ve, and -ve values of $K$, respectively. We use $log(K)$ scale for the +ve values of $K$ for better visualization. (c) Variation of the Fano factor ($f$) with $\langle N(T)\rangle$ for a range of values of $k_{-1}$.  The points depicting variations of $f$ with $\langle N(T)\rangle$  as $T$ is increased at a fixed value of $k_{-1}$ are connected with solid lines.  Increasing values of $k_{-1}$ are indicated by changing the shade of the symbols from dark to lighter shades of gray. $k_{-1}$ changes uniformly from $0.01$ to $0.208$ in an interval of $0.018$ for the feedback. For the linear model, $k_{-1}$ changes evenly from $0.0005$ to $0.0995$ in an interval of $0.009$.  All the other parameters for both the models are held fixed at the values given in (a). (d) Variation of $K$ with $\langle N(T)\rangle$ for the feedback and the linear model displayed using the same visualization scheme and parameters as in (c). The feedback and the linear models produced large +ve, and -ve values of $K$, respectively. We use $log(K)$ scale for the +ve values of $K$ for better visualization. }
\label{fig:revfanokurt}
\end{figure}

\vspace{10mm}
\pagebreak


\clearpage 

\begin{thebibliography}{11}

\bibitem{DAS}
J. Das, M. Ho, J. Zikherman, C. Govern, M. Yang, A. Weiss, A. K. Chakraborty, and J. P. Roose, Cell {\bf 136}, 337 (2009).

\bibitem{FERR1}
J. E. Ferrell Jr and E. M. Machleder, Science {\bf 280}, 895 (1998).

\bibitem{COLLINS}
T. S. Gardner, C. R. Cantor, and J. J. Collins, Nature {\bf 403}, 339 (2000).

\bibitem{GARDN}
C. Gardiner, {\it{Handbook of Stochastic Methods for Physics, Chemistry and Natural Sciences}}, (Springer-Verlag, Heidelberg, 2004).

\bibitem{DELBRUCK}
M. Delbruck, J. Chem. Phys. {\bf 8}, 120 (1940).

\bibitem{ARKIN}
H. H. McAdams and A. Arkin, Proc. Natl. Acad. Sci. USA {\bf 94}, 814 (1997).


\bibitem{WEINBERGER}
L. S. Weinberger and T. Shenk, PloS Biol. {\bf 5}, e9 (2007).


\bibitem{SIGGIA}
P. S. Swain, M. B. Elowitz, and E. D. Siggia, Proc. Natl. Acad. Sci. USA {\bf 99}, 12795 (2002).

\bibitem{GUMB}
E. J. Gumbel, {\it Statistics of Extremes}, (Dover Publications, Inc, New York, 2004).

\bibitem{MAJUM1}
S. N. Majumdar and A. Comtet, Phys. Rev. Lett. {\bf 92} 225501 (2004).

\bibitem{MAJUM2}
S. N. Majumdar, J. Randon-Furling, M. J. Kearney, and M. Yor, J. Phys. A {\bf 41} 365005 (2008).

\bibitem{KADAN}
L. P. Kadanoff, Chin. J. Phys. {\bf 29} 613 (1993).

\bibitem{GILLESPIE1}
J. H. Gillespie, Theor. Popul. Biol. {\bf 23}, 202 (1983).


\bibitem{CHAK}
A. Kosmerlj, A. K. Chakraborty, M. Kardar, and E. I. Shakhnovich, Phys. Rev. Lett. {\bf 103} 068103 (2009).

\bibitem{FINANCE}
P. Embrechts, C. Kluppelberg, and T. Mikosch, {\it Modelling Extremal Events for Insurance and Finance} (Springer, Berlin, 2004).

\bibitem{ORR}
H. A. Orr, Evolution {\bf 56}, 1317 (2002).

\bibitem{KARDAR}
M. Artomov, M. Kardar, and A. K. Chakraborty, J. Chem. Phys. {\bf 133}, 105101 (2010).

\bibitem{RACZ}
T. W. Burkhard, G. Gyorgyi, N. R. Moloney, and Z. Racz, Phys. Rev. E {\bf 76}, 041119 (2007).

\bibitem{REDNER}
S. Redner, {\it A Guide to First-Passage Processes}, (Cambridge University Press, Cambridge, 2001).

\bibitem{GOEL}
N. Goel and N. Ritcher-Dyn, {\it Stochastic Models in Biology}, (Academic Press, NewYork, 1974).

\bibitem{HONER}
J. Honerkamp, {\it Stochastic Dynamical Systems}, (VCH Publishers, New York, 1993).

\bibitem{DECAR}
L. T. DeCarlo, Psychol. Meth. {\bf 2}, 292 (1997).

\bibitem{GILLESPIE}
D. T. Gillespie, J. Chem. Phys. {\bf 81}, 2340 (1977).

\bibitem{MEZARD}
J-P. Bouchaud and M. Mezard, J. Phys. A {\bf 30} 7997 (1997).

\bibitem{KATZ}
R. W. Katz and B. G. Brown, Clim. Chan. {\bf 21}, 289 (1992).

\bibitem{SSC}
M. Lis, M. N. Artyomov, S. Devadas, A. K. Chakraborty, Bioinform. {\bf 25}, 2289(2009).

\bibitem{SUPPL}
See Supplementary Material  Document No.   $\,\,\,\,\,$  for additional details.

\end{thebibliography}
\end{document}